\documentclass[twocolumn]{aastex631}

\usepackage[utf8]{inputenc}
\usepackage{graphicx,psfrag}
\usepackage{mathrsfs}
\usepackage{amsmath,amsfonts,amssymb}
\usepackage{multirow} 
\usepackage{diagbox}
\usepackage{comment}
\usepackage{xcolor}
\usepackage{enumerate}
\usepackage{booktabs}
\usepackage{mathtools}
\usepackage{color,soul}

\usepackage{hyperref}
\hypersetup{
    colorlinks = true,
    linkcolor = {blue},
    citecolor = {blue},
    urlcolor = {blue},
    linkbordercolor = {white},
    }

\usepackage{color}
\definecolor{cyan}{rgb}{0,0.9,0.9}
\definecolor{orange}{rgb}{0.9,0.5,0}
\definecolor{magenta}{rgb}{1,0,1}
\definecolor{purple}{rgb}{0.8,0.4,0.8}
\definecolor{gray}{rgb}{0.8242,0.8242,0.8242}
\definecolor{mgreen}{rgb}{0.1,0.8,0.1}

\usepackage[normalem]{ulem}

\begin{document}




\title{What can we learn about the unstable equation-of-state branch from neutron-star mergers?}

\author{Maximiliano Ujevic}
\affiliation{Centro de Ci$\hat{e}$ncias Naturais e Humanas, Universidade Federal do ABC, 09210-170, Santo Andr{\'e}, 9210-170, SP, Brazil}

\author{Rahul \surname{Somasundaram}}
\affiliation{Theoretical Division, Los Alamos National Laboratory, Los Alamos, NM 87545, USA}
\affiliation{Department of Physics, Syracuse University, Syracuse, NY 13244, USA}

\author{Tim \surname{Dietrich}}
\affiliation{Institute for Physics and Astronomy, University of Potsdam, D-14476 Potsdam, Germany}
\affiliation{Max Planck Institute for Gravitational Physics (Albert Einstein Institute), Am M\"uhlenberg 1, Potsdam 14476, Germany}

\author{Jerome \surname{Margueron}}
\affiliation{Institut de Physique des 2 infinis de Lyon, CNRS/IN2P3, Universit\'e de Lyon, Universit\'e Claude Bernard Lyon 1, F-69622 Villeurbanne Cedex, France
}
\affiliation{International Research Laboratory on Nuclear Physics and Astrophysics, Michigan State University and CNRS, East Lansing, MI 48824, USA} 

\journalinfo{}

\author{Ingo \surname{Tews}}
\affiliation{Theoretical Division, Los Alamos National Laboratory, Los Alamos, NM 87545, USA}

\date{\today}

\reportnum{LA-UR-23-31683}

\begin{abstract} 
The Equation of State (EOS) of dense strongly-interacting matter can be probed by astrophysical observations of neutron stars (NS), such as X-ray detections of pulsars or the measurement of the tidal deformability of NSs during the inspiral stage of NS mergers. 
These observations constrain the EOS at most up to the density of the maximum-mass configuration, $n_\textrm{TOV}$, which is the highest density that can be explored by stable NSs for a given EOS. 
However, under the right circumstances, binary neutron star (BNS) mergers can create a postmerger remnant that explores densities above $n_\textrm{TOV}$. 
In this work, we explore whether the EOS above $n_\textrm{TOV}$ can be measured from gravitational-wave or electromagnetic observations of the postmerger remnant. 
We perform a total of twenty-five numerical-relativity simulations of BNS mergers for a range of EOSs and find no case in which different descriptions of the matter above $n_{\rm TOV}$ have a detectable impact on postmerger observables.
Hence, we conclude that the EOS above $n_\textrm{TOV}$ can likely not be probed through BNS merger observations for the current and next generation of detectors. 
\end{abstract}

\section{Introduction}
\label{sec:intro}

The Equation of State (EOS) of dense matter is a fundamental relation in nuclear (astro)physics.
It connects the properties of strong interactions among the relevant microscopic degrees of freedom, described by quantum Chromodynamics (QCD), to the global properties of neutron stars (NSs), such as their masses, radii, and tidal deformabilities. 
In recent years, astrophysical measurements of NS properties from X-ray~\citep{Miller:2019cac,Riley:2019yda,Miller:2021qha,Riley:2021pdl}, radio~\citep{Demorest:2010bx,Antoniadis:2013pzd,NANOGrav:2019jur,Fonseca:2021wxt}, and gravitational wave (GW) observations of the inspiral phase of binary neutron star (BNS) mergers~\citep{LIGOScientific:2017qsa,LIGOScientific:2018cki,LIGOScientific:2018hze,Abbott:2020uma} have provided a wealth of information on the EOS, see e.g.,~\citep{Margalit:2017dij,Bauswein:2017vtn,Annala:2017llu,Most:2018hfd,Ruiz:2017due,Radice:2018ozg,Landry:2018prl,Capano:2019eae,Raaijmakers:2019dks,Dietrich:2020efo,Essick:2020flb,Raaijmakers:2021uju,Huth:2021bsp,Pang:2022rzc,Ghosh:2022lam}. 
While these observations already led to exciting results, the era of high-precision astrophysical measurements of the EOS is yet to come, with the next generation of GW detectors being planned in the United States~\citep{Reitze:2019iox,Evans:2021gyd} and Europe~\citep{Punturo:2010zz,Branchesi:2023mws}, and improved large-area X-ray timing telescopes anticipated in the future, e.g.,\citep{Mushotzky:2019lpm,2019JATIS...5b1001G,eXTP:2018anb}.

The EOS of dense matter links the pressure $p$, energy density $\epsilon$, and the number density $n$, spanning from very dilute matter up to asymptotically large densities where the dynamics of QCD become perturbative, $n ~\sim 40 n_\textrm{sat}$ with $n_\textrm{sat}$ being the nuclear saturation density. 
However, stable NSs probe the EOS only up to the central density of the maximum-mass configuration, which we call the TOV density $n_\textrm{TOV}$ in the following. 
This limit results from the General Relativity (GR) structure equations for NSs, the TOV equations, and depends on the EOS for $n \leq n_\textrm{TOV}$. 
While the exact value of $n_\textrm{TOV}$ is unknown, present astrophysical observations place it at $n_\textrm{TOV} \sim 5 - 8 n_\textrm{sat}$~\citep{Pang:2022rzc}. 

\begin{figure}[t]
\centering\includegraphics[width=\columnwidth]{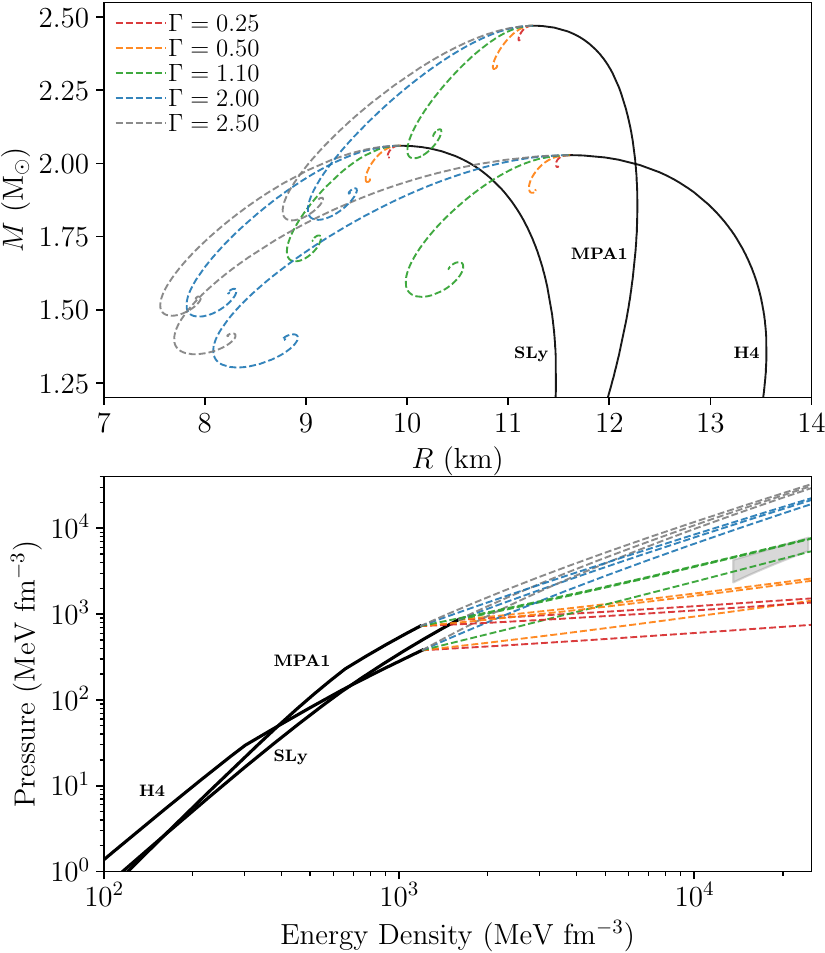}
    \caption{{\it Upper panel}: Mass-radius curves for the three stable-branch EoSs employed in this work. 
    The EoSs have different compactness and maximum masses, covering a considerable portion of the NS parameter space. With dashed lines and different colors, we represent the various extensions for the unstable branches. 
    {\it Bottom panel}: Pressure vs. energy density for the EoSs and extensions used in this work. The grey area shows the perturbative QCD domain.}
    \label{fig:EOS}
\end{figure}

Consequently, observations of pulsars or NS inspirals cannot probe the EOS above $n_\textrm{TOV}$.
Nevertheless, the EOS above $n_\textrm{TOV}$ is of great interest.
Its properties might affect $n_\textrm{TOV}$ because it is possible that a strong softening, such as a phase transition slightly above $n_\textrm{TOV}$, destabilizes NSs, and hence, setting the properties of the maximum-mass configuration to begin with.
Furthermore, in the postmerger phase of BNS collisions, densities in the remnant might exceed $n_\textrm{TOV}$ as differential rotation aids stability of the system.
Then, the EOS above $n_\textrm{TOV}$ might be relevant to describe postmerger physics, such as the postmerger GW signal or the mass of the ejecta, and hence, the kilonova lightcurves. 

In this Letter, we systematically investigate the possibility of measuring the EOS above $n_\textrm{TOV}$ with observations of the postmerger remnant of BNS mergers.
We perform a simulation campaign with 15 EOSs, considering different behavior for the stable and unstable branches, see Fig.~\ref{fig:EOS}.
We examine the GWs emitted by the remnant, the maximum density explored in the simulation, and the ejecta masses in detail, and find that, in all considered cases, there is no detectable impact of the EOS above $n_\textrm{TOV}$ on any observable related to the postmerger phase. 
Our findings establish $n_\textrm{TOV}$ as a firm limit up to which physicists can hope to probe the EOS with astrophysical data.
They further indicate that, given the expected uncertainties in the next decades, accurate knowledge of the cold and hot EOS below $n_\textrm{TOV}$ is sufficient for the prediction of postmerger observables such as the ejecta mass and the GW waveform.

\section{Setup}
\label{sec:setup}

\subsection{Equations of State}

In this study, we perform a simulation campaign with 15 different EOSs.
These EOSs explore different behavior in the stable and unstable NS branches. 
In particular, we employ the SLy~\citep{Douchin:2001sv}, MPA1~\citep{Muther:1987xaa}, and H4~\citep{Lackey:2005tk} EOSs up to $n_\textrm{TOV}$.
These EOSs are implemented in the form of their polytropic approximations from \cite{Read:2008iy}.
All three EOSs explore different NS radii, ranging from $\sim11.5-13.5$~km for a typical $1.4~M_{\rm sol}$ NS.
The SLy and H4 EOS have similar maximum masses of $\sim 2~M_{\rm sol}$, while the MPA1 EOS has a maximum mass of $\sim 2.5~M_{\rm sol}$.
Hence, these three EOS represent different possible behaviors in the stable NS branch, see Fig.~\ref{fig:EOS}, and enable us to test the impact of this part of the EOS.

To probe the impact of the unstable branch, we attach another polytropic segment at $n_{\rm TOV}$ to each EOS. 
For this segment, we vary the polytropic index $\Gamma$ between 0.25 and 2.5, exploring different stiffness in the unstable branch up to high densities, see Fig.~\ref{fig:EOS}.
At the highest densities, $n \sim 40 n_\textrm{sat}$, the EOS can be constrained by perturbative QCD (pQCD) calculations~\cite{Gorda:2018gpy,Gorda:2021znl,Gorda:2023mkk}.
Our EOSs with $\Gamma=1.1$ are consistent with results from pQCD in the $p-\epsilon$ plane, while the other extensions are either too soft or too stiff to be in agreement with pQCD. 

\cite{Komoltsev:2021jzg} showed that pQCD calculations can also constrain the EOS at lower densities by imposing causality and thermodynamic consistency of the EOS with high-density pQCD calculations. 
Using this approach, \cite{Gorda:2022jvk} outlined that pQCD impacts the EoS of stable NSs while \cite{Somasundaram:2022ztm} found that the integral pQCD constraints have a marginal
impact on the NS EoS selection once nuclear physics and astrophysical sources of information are accounted for.
However, pQCD impacts the EOS in the unstable branch above $n_\textrm{TOV}$, see Fig.~\ref{fig:EOS}. 
The present work will allow us to quantify more generally if densities above $n_\textrm{TOV}$ can be measured through astrophysical observations.

\subsection{Numerical Relativity Simulations}

For the construction of our initial configurations, we are using the pseudo-spectral SGRID code~\citep{Tichy:2009yr,Tichy:2012rp,Dietrich:2015pxa,Tichy:2019ouu}. 
SGRID uses surface fitting coordinates to solve the Einstein Equations following the extended conformal thin sandwich (XCTS) formulation~\citep{York:1998hy}. 

For the dynamical simulations, we use the finite-differencing numerical-relativity (NR) code BAM~\citep{Bruegmann:2006ulg,Thierfelder:2011yi,Dietrich:2015iva,Bernuzzi:2016pie,Dietrich:2018phi}. 
We use the Z4c formulation for the spacetime evolution~\citep{Bernuzzi:2009ex,Hilditch:2012fp}, the moving puncture gauge~\citep{Bona:1994a,Alcubierre:2002kk,vanMeter:2006vi}, and the Valencia formulation for general-relativistic hydrodynamics~\citep{Marti:1991wi,Banyuls:1997zz,Anton:2005gi} together with high-resolution shock-capturing techniques for the description of matter variables. For the BAM simulations, we augment the cold EOSs described before with a $\Gamma$-law EOS using $\Gamma_{\rm thermal}=1.75$ to incorporate thermal effects~\cite{Bauswein:2010dn}.

To ensure that we can resolve the relevant length scales, i.e., the far-field region in which GWs are extracted, but also the matter flow inside the stars, BAM uses a box-in-box mesh refinement that automatically can follow the movement of the stars. 
A Berger-Collela refinement strategy ensures stable and accurate simulations~\citep{Dietrich:2015iva}.
For this work, we employ a total of seven refinement levels and a resolution of approximately 108~m in the finest one.

\section{Results}
\label{sec:results}

\begin{figure*}[t]
    \centering
    \includegraphics[width=\textwidth]{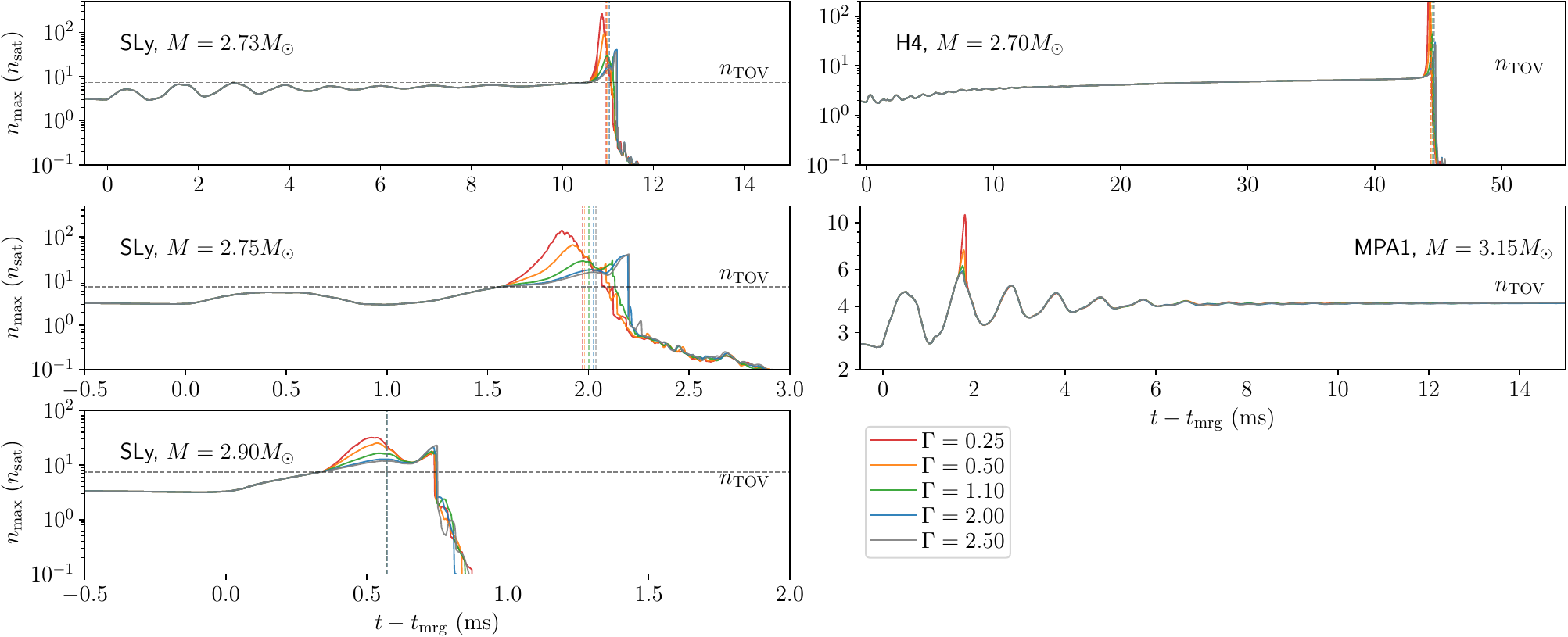}
    \caption{Time evolution of the maximum density, $n_{\rm max}$, for all 25 simulations. We mark the maximum density present inside a non-rotating NS at the TOV mass ($n_{\rm TOV}$) with a horizontal dashed line.
    In most cases, once this density is reached, the remnant collapses quickly into a black hole. 
    However, our MPA1 $M=3.15M_\odot$ is a clear exception where, for a short amount of time, the central density exceeds $n_{\rm TOV}$, but the mass contained in the central region of the remnant is not large enough to trigger black hole formation. 
    When available, we plot as vertical lines the time when the apparent horizon is first found in our simulation.
    }
    \label{fig:rho_max}
\end{figure*}

\begin{figure*}[t]
    \centering
    \includegraphics[width=\textwidth]{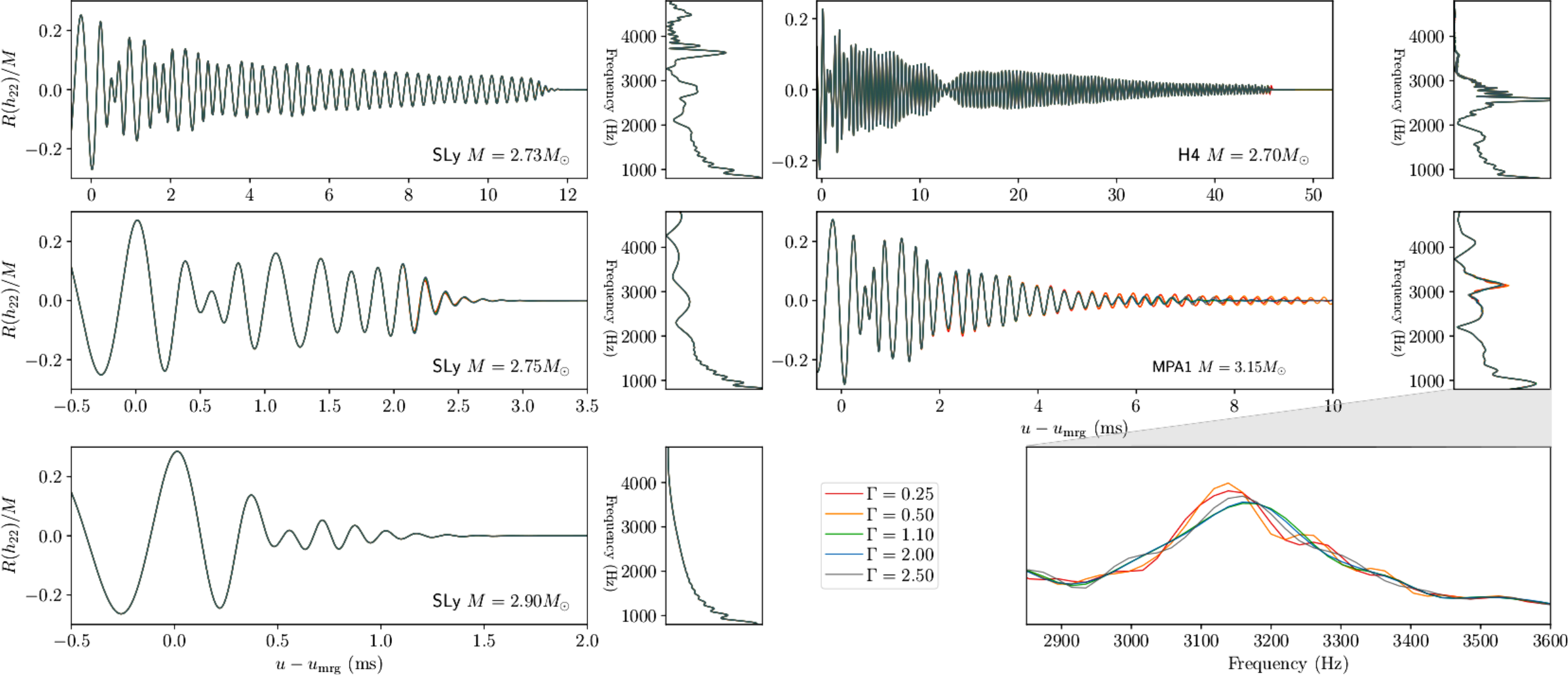}
    \caption{GW strain for all considered scenarios. We show the real part of the (2,2)-mode of the GW signal in the main panels and the corresponding spectrograms from the entire simulation in the smaller panels. 
    We find a visible difference between the simulations with different high-density EOS only for one scenario, the MPA1-like simulations, and present an inset of the spectrogram in the bottom right panel of the figure. 
    While there is a shift in the peak frequency of about 30~Hz, the widths of the individual peaks are significantly larger. 
    This will make it almost impossible to distinguish between the different EOSs, even with future-generation GW detectors.
    }
    \label{fig:waves}
\end{figure*}

\begin{figure*}[t]
    \centering
    \includegraphics[width=\textwidth]{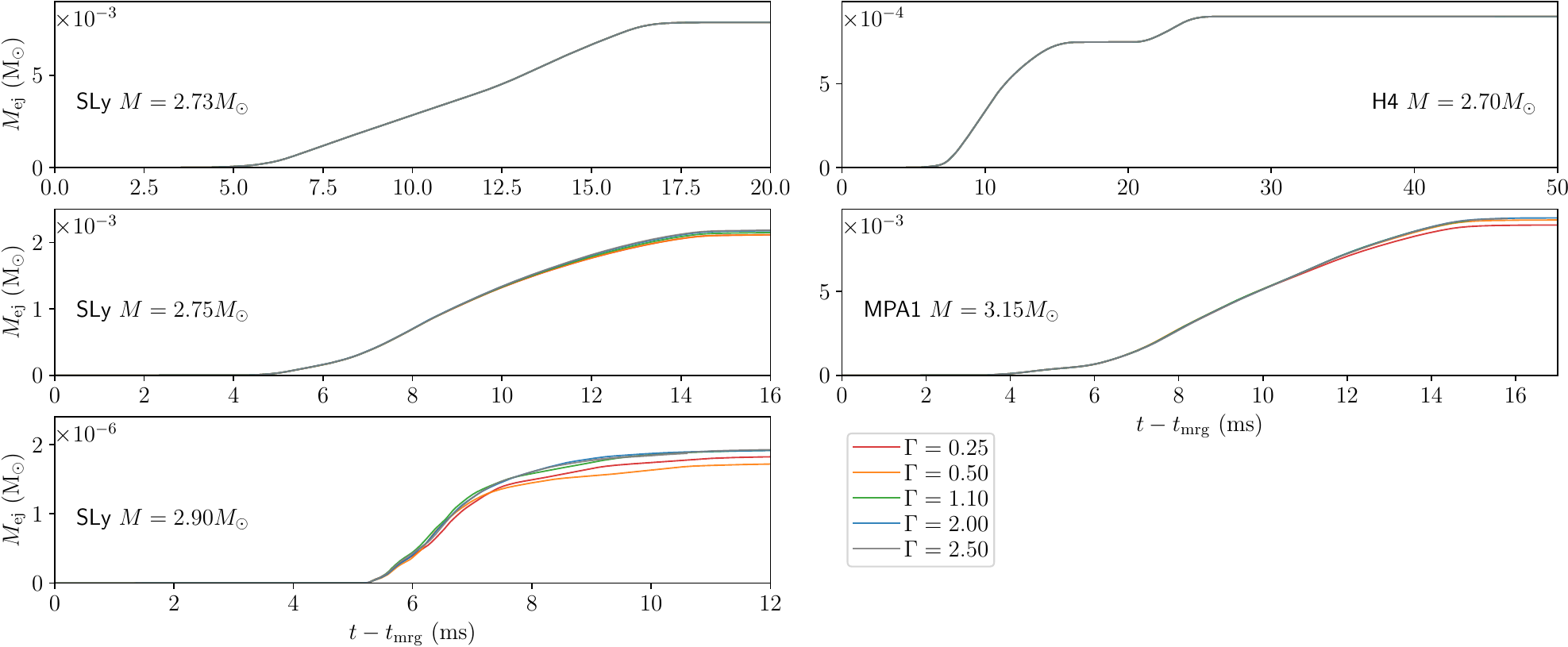}
    \caption{The ejecta mass for all considered EOS scenarios as a function of time. 
    The ejecta masses are extracted on a sphere with a coordinate radius of about $600 {\rm km}$.}
    \label{fig:ejecta}
\end{figure*}

To quantify the possible impact of the EOS in the unstable branch above $n_{\rm TOV}$, we will discuss the evolution of the maximum density and the black hole formation (Fig.~\ref{fig:rho_max}), outline possible impacts on the postmerger GW signal (Fig.~\ref{fig:waves}), and report on the released ejecta mass (Fig.~\ref{fig:ejecta}). 

\subsection{Density Evolution and Black Hole Formation}

For our SLy-EOS simulations, we consider three different total masses: $2.73M_\odot$, $2.75 M_\odot$, and $2.90M_\odot$; left panels of Fig.~\ref{fig:rho_max}.
Considering the $2.73M_\odot$ setup, we find a difference in the collapse time, i.e., the time between merger and formation of a black hole, of about ${\cal O}(100 \mu s)$ for the different EOS extensions.
This time difference is too small to be detectable in upcoming BNS merger observations. 
Overall, our findings suggest that a black hole forms very quickly once the central density reaches $n_{\rm TOV}$.
Once this critical density is reached, higher densities are present during the ongoing simulation, and the evolution differs between the different setups.  
Surprisingly, we find that the densities peak before the formation of an apparent horizon, reported as vertical dashed lines in Fig.~\ref{fig:rho_max}. 
We note that there is a high chance that the apparent horizon formed earlier but was not tracked by the employed algorithm immediately, i.e., the dashed lines report an upper bound on the time when a black hole forms. 
We believe that, in practice, the highest densities are reached within a region either located inside the formed apparent horizon or in a region where an apparent horizon forms very quickly. 
The densities drop once they reach the singularity since (i) the employed shift conditions excise the central singularity, and (ii) once the fluxes become infinite (or extremely large), the matter is removed from our grid~\citep{Thierfelder:2010dv}. In general, we find that this drop in the density happens slightly later for larger values of $\Gamma$, likely due to the larger pressure for larger values of $\Gamma$. 

For the $2.75 M_\odot$ scenario, there is a small shift in the collapse time and a slight difference in the simulation once $n_{\rm TOV}$ is exceeded. 
However, the time difference is again ${\cal O}(100 \mu s)$.
Finally, for the $2.90M_\odot$ simulation, we find that the collapse to a black hole happens within $1 \rm ms$ after the merger for all setups almost independently of the EOS extension, i.e., there is no measurable difference between the individual scenarios. 

The previous simulations all use the same stable-branch EOS.
To explore the impact of the stable branch, we also consider two other EOSs. 
For the H4 setup, we chose a total mass of $2.70 M_\odot$ for our simulation to ensure that the merger remnant reaches densities above $n_{\rm TOV}$. 
In this case, the remnant survives for more than $40$~ms before a black hole forms.
We find that although the collapse happens much later than for the SLy scenarios, there is almost no difference between different values of $\Gamma$ above $n_{\rm TOV}$. 

Finally, we studied the MPA1 EOS with the highest maximum mass. 
For this setup, we carefully fine-tuned the masses performing different (low resolution) simulations and settled on a total mass of $3.15 M_\odot$, which shows overall the largest imprint of the unstable branch. 
For this setup, we did not observe a collapse to a black hole during the time of our simulation. 
However, we found that the star surpasses $n_{\rm TOV}$ about 2~ms after the merger but did not form a black hole (increasing the total mass led to black hole formation).
On the contrary, the star's central density decreases again, continuing the density oscillation. 
This is of particular interest since it implies that for a short amount of time, higher densities than $n_{\rm TOV}$ were probed. 
Nevertheless, this effect does not lead to detectable differences. 

These results are quite interesting. 
Long lifetimes lead to remnants that explore central densities well below $n_{\rm TOV}$, minimizing the impact of the unstable branch. 
When increasing the mass, and reducing the lifetime of the remnant, appearing differences become undetectably short. 
From our investigations, we find that there seems to be no ``sweet spot'' in between where the unstable branch has a discernable impact.

\subsection{Gravitational Waves}

For the computation of the predicted GW signals, we are extracting the Newman-Penrose scalar $\Psi_4$ on an extraction sphere of radius about 1200~km around the origin of our computational domain and then compute the GW strain 
$h(t) = - \int_{-\infty}^t  \int_{-\infty}^{t'} \Psi_4 \text{d}t'' \text{d}t'$ following~\cite{Reisswig:2010di}. 
We show the GW strain and the corresponding power spectrum from the entire simulation in Fig.~\ref{fig:waves}. 
We find a perfect agreement for all SLy-like and H4-like simulations, i.e., no noticeable imprint of the unstable EOS branch on the GW signals is visible.  
Only for the MPA1 EOS do we find a slight difference in the GW signal due to a small shift in the oscillation of the remnant around $2~{\rm ms}$ after the merger; see Fig.~\ref{fig:rho_max}). 
This difference results in a shift of about $30~{\rm Hz}$ in the GW peak frequency between the most extreme scenarios, for $\Gamma=0.25$ and $\Gamma=2.50$. 
Previous studies revealed measurement uncertainties of the peak frequency of about $50$ to $100~{\rm Hz}$ for close BNS mergers for which the postmerger signal-to-noise ratio would be of the order $\mathcal{O}(10)$, see e.g., \cite{Clark:2015zxa, Rezzolla:2016nxn,Breschi:2022ens} or \cite{Branchesi:2023mws}. 
We also point out that it would be extremely difficult to this a shift due to the unstable-branch EOS from other effects, such as magnetic fields or neutrino radiation. 
For these reasons, we do not expect that similar differences would be detectable in the postmerger spectrum in future GW observations. 

Overall, considering the need to fine-tune the system parameters to find imprints of the unstable EOS branch on the emitted GW signal and the minimal frequency shifts in the postmerger peak frequency, we conclude that it is unlikely that GW information can be used to access the unstable EOS branch. 

\subsection{Ejecta}

In addition to the emitted GW signals, observations of electromagnetic counterparts might help us to distinguish between different EOSs in the unstable branch.
Given that the brightness of the electromagnetic counterparts is mainly proportional to the amount of ejected material, we will use the mass of ejected matter as an indicator for the possibility to differentiate between the different scenarios considered in this work. 
For this reason, we track outward-flowing, unbound material (using the geodesic criterion, i.e., $u_t < -1$). 
The extraction sphere is centered around the origin and has a radius of about 600~km. 

For our SLy-like simulations, we generally find a decreasing ejecta mass with an increasing total mass. 
However, we do not find a measurable signal for different extensions of the EOS above $n_{\rm TOV}$. 
Similarly, also for the H4 and MPA1 setups, the differences in ejecta masses for the various EOS extensions are below $10^{-4}M_{\odot}$.
This difference is well below the measurement uncertainties of current electromagnetic observations of a kilonova and its modeling uncertainties due to incomplete knowledge of opacities, heating rates, or thermal efficiencies~\citep{Dietrich:2020efo,Heinzel:2020qlt,Fryer:2023pew}. 
Therefore, we can conclude that for all employed scenarios, we will not be able to access the EOS above $n_{\rm TOV}$ through the observations of electromagnetic counterparts.

\section{Summary and Conclusions}
\label{sec:conclusion}

Given previous studies about the possibility of measuring the high-density EOS from neutron star mergers, e.g.,~\cite{Bauswein:2018bma,Most:2018eaw,Figura:2020fkj}, we extend this question in this letter and have have investigated if binary neutron star mergers could provide us with meaningful information about the unstable branch of the EOS, i.e., the EOS at densities beyond the maximum density of a non-rotating single neutron star. 
For this reason, we have performed numerical-relativity simulations for three different EOSs (SLy, H4, MPA1) to describe dense matter up to $n_{\rm TOV}$.
Above this density, we have used five different extensions per EOS to investigate if there is an effect of the unstable branch on observable quantities during the binary NS merger. 
To increase the sample size of our study, we have used five different total masses, selected in such a way as to ensure that $n_{\rm TOV}$ is reached during the postmerger evolution. 

During our simulations, we found no noticeable effect of the EOS in the unstable branch in any of our simulations.
Indeed, the collapse time, i.e., the time between the merger and the black hole formation, only changes by a few hundred microseconds, and the emitted GW signal and the amount of ejected material seem to be independent of the particular extension used above $n_{\rm TOV}$. 
We point out that some of these extensions ($\Gamma =2.50$) are extreme and become even non-causal in some parts of the simulation. 
Based on these observations and given current and expected future observational uncertainties, we suggest that it is unrealistic to expect that binary NS mergers will allow us to probe densities beyond those probed in high-mass pulsars. 
Certainly, we cannot rule out that there might be a particular combination of source parameters or a phase transition leading to sudden density jumps that might be observable.
However, given our set of simulations, the attempt to fine-tune parameters to find such a postmerger difference, and the low signal-to-noise ratio that most binary NS observations are expected to have, it seems unlikely that binary NS mergers can be used to test densities above $n_{\rm TOV}$.

\vspace{0.3cm}

Co-funded by the European Union (ERC, SMArt, 101076369). Views and opinions expressed are those of the authors only and do not necessarily reflect those of the European Union or the European Research Council. Neither the European Union nor the granting authority can be held responsible for them.
R.S. acknowledges support from the Nuclear Physics from Multi-Messenger Mergers (NP3M) Focused Research Hub which is funded by the National Science Foundation under Grant Number 21-16686, and by the Laboratory Directed Research and Development program of Los Alamos National Laboratory under project number 20220541ECR.
I.T. was supported by the U.S. Department of Energy, Office of Science, Office of Nuclear Physics, under contract No.~DE-AC52-06NA25396, by the Laboratory Directed Research and Development program of Los Alamos National Laboratory under project numbers 20220541ECR and 20230315ER, and by the U.S. Department of Energy, Office of Science, Office of Advanced Scientific Computing Research, Scientific Discovery through Advanced Computing (SciDAC) NUCLEI program.


\bibliography{refs.bib}

\begin{thebibliography}{}
\expandafter\ifx\csname natexlab\endcsname\relax\def\natexlab#1{#1}\fi
\providecommand{\url}[1]{\href{#1}{#1}}
\providecommand{\dodoi}[1]{doi:~\href{http://doi.org/#1}{\nolinkurl{#1}}}
\providecommand{\doeprint}[1]{\href{http://ascl.net/#1}{\nolinkurl{http://ascl.net/#1}}}
\providecommand{\doarXiv}[1]{\href{https://arxiv.org/abs/#1}{\nolinkurl{https://arxiv.org/abs/#1}}}

\bibitem[{Abbott {et~al.}(2020)}]{Abbott:2020uma}
Abbott, B., {et~al.} 2020, Astrophys. J. Lett., 892, L3,
  \dodoi{10.3847/2041-8213/ab75f5}

\bibitem[{Abbott {et~al.}(2017)}]{LIGOScientific:2017qsa}
Abbott, B.~P., {et~al.} 2017, Phys. Rev. Lett., 119, 161101,
  \dodoi{10.1103/PhysRevLett.119.161101}

\bibitem[{Abbott {et~al.}(2018)}]{LIGOScientific:2018cki}
---. 2018, Phys. Rev. Lett., 121, 161101,
  \dodoi{10.1103/PhysRevLett.121.161101}

\bibitem[{Abbott {et~al.}(2019)}]{LIGOScientific:2018hze}
---. 2019, Phys. Rev. X, 9, 011001, \dodoi{10.1103/PhysRevX.9.011001}

\bibitem[{Alcubierre {et~al.}(2003)Alcubierre, Br{\"u}gmann, Diener, Koppitz,
  Pollney, {et~al.}}]{Alcubierre:2002kk}
Alcubierre, M., Br{\"u}gmann, B., Diener, P., {et~al.} 2003, Phys.Rev., D67,
  084023, \dodoi{10.1103/PhysRevD.67.084023}

\bibitem[{Annala {et~al.}(2018)Annala, Gorda, Kurkela, \&
  Vuorinen}]{Annala:2017llu}
Annala, E., Gorda, T., Kurkela, A., \& Vuorinen, A. 2018, Phys. Rev. Lett.,
  120, 172703, \dodoi{10.1103/PhysRevLett.120.172703}

\bibitem[{Anton {et~al.}(2006)Anton, Zanotti, Miralles, Marti, Ibanez,
  {et~al.}}]{Anton:2005gi}
Anton, L., Zanotti, O., Miralles, J.~A., {et~al.} 2006, Astrophys.J., 637, 296,
  \dodoi{10.1086/498238}

\bibitem[{Antoniadis {et~al.}(2013)Antoniadis, Freire, Wex, Tauris, Lynch,
  {et~al.}}]{Antoniadis:2013pzd}
Antoniadis, J., Freire, P.~C., Wex, N., {et~al.} 2013, Science, 340, 6131,
  \dodoi{10.1126/science.1233232}

\bibitem[{Banyuls {et~al.}(1997)Banyuls, Font, Ibanez, Marti, \&
  Miralles}]{Banyuls:1997zz}
Banyuls, F., Font, J.~A., Ibanez, J. M.~A., Marti, J. M.~A., \& Miralles, J.~A.
  1997, Astrophys. J., 476, 221

\bibitem[{Bauswein {et~al.}(2019)Bauswein, Bastian, Blaschke, Chatziioannou,
  Clark, Fischer, \& Oertel}]{Bauswein:2018bma}
Bauswein, A., Bastian, N.-U.~F., Blaschke, D.~B., {et~al.} 2019, Phys. Rev.
  Lett., 122, 061102, \dodoi{10.1103/PhysRevLett.122.061102}

\bibitem[{Bauswein {et~al.}(2010)Bauswein, Janka, \&
  Oechslin}]{Bauswein:2010dn}
Bauswein, A., Janka, H.-T., \& Oechslin, R. 2010, Phys.Rev., D82, 084043,
  \dodoi{10.1103/PhysRevD.82.084043}

\bibitem[{Bauswein {et~al.}(2017)Bauswein, Just, Janka, \&
  Stergioulas}]{Bauswein:2017vtn}
Bauswein, A., Just, O., Janka, H.-T., \& Stergioulas, N. 2017, Astrophys. J.,
  850, L34, \dodoi{10.3847/2041-8213/aa9994}

\bibitem[{Bernuzzi \& Dietrich(2016)}]{Bernuzzi:2016pie}
Bernuzzi, S., \& Dietrich, T. 2016, Phys. Rev., D94, 064062,
  \dodoi{10.1103/PhysRevD.94.064062}

\bibitem[{Bernuzzi \& Hilditch(2010)}]{Bernuzzi:2009ex}
Bernuzzi, S., \& Hilditch, D. 2010, Phys. Rev., D81, 084003,
  \dodoi{10.1103/PhysRevD.81.084003}

\bibitem[{Bona {et~al.}(1996)Bona, Mass{\'o}, Stela, \& Seidel}]{Bona:1994a}
Bona, C., Mass{\'o}, J., Stela, J., \& Seidel, E. 1996, in The Seventh {M}arcel
  {G}rossmann Meeting: On Recent Developments in Theoretical and Experimental
  General Relativity, Gravitation, and Relativistic Field Theories, ed. R.~T.
  Jantzen, G.~M. Keiser, \& R.~Ruffini (Singapore: World {S}cientific)

\bibitem[{Branchesi {et~al.}(2023)}]{Branchesi:2023mws}
Branchesi, M., {et~al.} 2023, JCAP, 07, 068,
  \dodoi{10.1088/1475-7516/2023/07/068}

\bibitem[{Breschi {et~al.}(2022)Breschi, Gamba, Borhanian, Carullo, \&
  Bernuzzi}]{Breschi:2022ens}
Breschi, M., Gamba, R., Borhanian, S., Carullo, G., \& Bernuzzi, S. 2022.
\newblock \doarXiv{2205.09979}

\bibitem[{Bruegmann {et~al.}(2008)Bruegmann, Gonzalez, Hannam, Husa, Sperhake,
  \& Tichy}]{Bruegmann:2006ulg}
Bruegmann, B., Gonzalez, J.~A., Hannam, M., {et~al.} 2008, Phys. Rev. D, 77,
  024027, \dodoi{10.1103/PhysRevD.77.024027}

\bibitem[{Capano {et~al.}(2020)Capano, Tews, Brown, Margalit, De, Kumar, Brown,
  Krishnan, \& Reddy}]{Capano:2019eae}
Capano, C.~D., Tews, I., Brown, S.~M., {et~al.} 2020, Nature Astron., 4, 625,
  \dodoi{10.1038/s41550-020-1014-6}

\bibitem[{Clark {et~al.}(2016)Clark, Bauswein, Stergioulas, \&
  Shoemaker}]{Clark:2015zxa}
Clark, J.~A., Bauswein, A., Stergioulas, N., \& Shoemaker, D. 2016, Class.
  Quant. Grav., 33, 085003, \dodoi{10.1088/0264-9381/33/8/085003}

\bibitem[{Cromartie {et~al.}(2019)}]{NANOGrav:2019jur}
Cromartie, H.~T., {et~al.} 2019, Nature Astron., 4, 72,
  \dodoi{10.1038/s41550-019-0880-2}

\bibitem[{Demorest {et~al.}(2010)Demorest, Pennucci, Ransom, Roberts, \&
  Hessels}]{Demorest:2010bx}
Demorest, P., Pennucci, T., Ransom, S., Roberts, M., \& Hessels, J. 2010,
  Nature, 467, 1081, \dodoi{10.1038/nature09466}

\bibitem[{Dietrich {et~al.}(2015{\natexlab{a}})Dietrich, Bernuzzi, Ujevic, \&
  Br{\"u}gmann}]{Dietrich:2015iva}
Dietrich, T., Bernuzzi, S., Ujevic, M., \& Br{\"u}gmann, B. 2015{\natexlab{a}},
  Phys. Rev., D91, 124041, \dodoi{10.1103/PhysRevD.91.124041}

\bibitem[{Dietrich {et~al.}(2020)Dietrich, Coughlin, Pang, Bulla, Heinzel,
  Issa, Tews, \& Antier}]{Dietrich:2020efo}
Dietrich, T., Coughlin, M.~W., Pang, P. T.~H., {et~al.} 2020, Science, 370,
  1450, \dodoi{10.1126/science.abb4317}

\bibitem[{Dietrich {et~al.}(2015{\natexlab{b}})Dietrich, Moldenhauer,
  Johnson-McDaniel, Bernuzzi, Markakis, Br{\"u}gmann, \&
  Tichy}]{Dietrich:2015pxa}
Dietrich, T., Moldenhauer, N., Johnson-McDaniel, N.~K., {et~al.}
  2015{\natexlab{b}}, Phys. Rev., D92, 124007,
  \dodoi{10.1103/PhysRevD.92.124007}

\bibitem[{Dietrich {et~al.}(2018)Dietrich, Radice, Bernuzzi, Zappa, Perego,
  Brügmann, Chaurasia, Dudi, Tichy, \& Ujevic}]{Dietrich:2018phi}
Dietrich, T., Radice, D., Bernuzzi, S., {et~al.} 2018, Class. Quant. Grav., 35,
  24LT01, \dodoi{10.1088/1361-6382/aaebc0}

\bibitem[{Douchin \& Haensel(2001)}]{Douchin:2001sv}
Douchin, F., \& Haensel, P. 2001, Astron. Astrophys., 380, 151

\bibitem[{Essick {et~al.}(2020)Essick, Tews, Landry, Reddy, \&
  Holz}]{Essick:2020flb}
Essick, R., Tews, I., Landry, P., Reddy, S., \& Holz, D.~E. 2020, Phys. Rev. C,
  102, 055803, \dodoi{10.1103/PhysRevC.102.055803}

\bibitem[{Evans \& others (Cosmic Explorer~Consortium)(2021)}]{Evans:2021gyd}
Evans, M., \& others (Cosmic Explorer~Consortium). 2021, arXiv:2109.09882

\bibitem[{Figura {et~al.}(2020)Figura, Lu, Burgio, Li, \&
  Schulze}]{Figura:2020fkj}
Figura, A., Lu, J.~J., Burgio, G.~F., Li, Z.~H., \& Schulze, H.~J. 2020, Phys.
  Rev. D, 102, 043006, \dodoi{10.1103/PhysRevD.102.043006}

\bibitem[{Fonseca {et~al.}(2021)}]{Fonseca:2021wxt}
Fonseca, E., {et~al.} 2021, Astrophys. J. Lett., 915, L12,
  \dodoi{10.3847/2041-8213/ac03b8}

\bibitem[{Fryer {et~al.}(2023)Fryer, Fontes, Korobkin, Mumpower, Wollaeger,
  Holmbeck, \& O\textquoteright{}Shaugnessy}]{Fryer:2023pew}
Fryer, C.~L., Fontes, C.~J., Korobkin, O., {et~al.} 2023, in {16th Marcel
  Grossmann Meeting on~Recent Developments in Theoretical and Experimental
  General Relativity, Astrophysics and Relativistic Field Theories},
  \dodoi{10.1142/9789811269776_0112}

\bibitem[{{Gaskin} {et~al.}(2019){Gaskin}, {Swartz}, {Vikhlinin}, {{\"O}zel},
  {Gelmis}, {Arenberg}, {Bandler}, {Bautz}, {Civitani}, {Dominguez}, {Eckart},
  {Falcone}, {Figueroa-Feliciano}, {Freeman}, {G{\"u}nther}, {Havey},
  {Heilmann}, {Kilaru}, {Kraft}, {McCarley}, {McEntaffer}, {Pareschi},
  {Purcell}, {Reid}, {Schattenburg}, {Schwartz}, {Schwartz}, {Tananbaum},
  {Tremblay}, {Zhang}, \& {Zuhone}}]{2019JATIS...5b1001G}
{Gaskin}, J.~A., {Swartz}, D.~A., {Vikhlinin}, A., {et~al.} 2019, Journal of
  Astronomical Telescopes, Instruments, and Systems, 5, 021001,
  \dodoi{10.1117/1.JATIS.5.2.021001}

\bibitem[{Ghosh {et~al.}(2022)Ghosh, Pradhan, Chatterjee, \&
  Schaffner-Bielich}]{Ghosh:2022lam}
Ghosh, S., Pradhan, B.~K., Chatterjee, D., \& Schaffner-Bielich, J. 2022,
  Front. Astron. Space Sci., 9, 864294, \dodoi{10.3389/fspas.2022.864294}

\bibitem[{Gorda {et~al.}(2023{\natexlab{a}})Gorda, Komoltsev, \&
  Kurkela}]{Gorda:2022jvk}
Gorda, T., Komoltsev, O., \& Kurkela, A. 2023{\natexlab{a}}, Astrophys. J.,
  950, 107, \dodoi{10.3847/1538-4357/acce3a}

\bibitem[{Gorda {et~al.}(2021)Gorda, Kurkela, Paatelainen, S\"appi, \&
  Vuorinen}]{Gorda:2021znl}
Gorda, T., Kurkela, A., Paatelainen, R., S\"appi, S., \& Vuorinen, A. 2021,
  Phys. Rev. Lett., 127, 162003, \dodoi{10.1103/PhysRevLett.127.162003}

\bibitem[{Gorda {et~al.}(2018)Gorda, Kurkela, Romatschke, S\"appi, \&
  Vuorinen}]{Gorda:2018gpy}
Gorda, T., Kurkela, A., Romatschke, P., S\"appi, S., \& Vuorinen, A. 2018,
  Phys. Rev. Lett., 121, 202701, \dodoi{10.1103/PhysRevLett.121.202701}

\bibitem[{Gorda {et~al.}(2023{\natexlab{b}})Gorda, Paatelainen, S\"appi, \&
  Sepp\"anen}]{Gorda:2023mkk}
Gorda, T., Paatelainen, R., S\"appi, S., \& Sepp\"anen, K. 2023{\natexlab{b}}.
\newblock \doarXiv{2307.08734}

\bibitem[{Heinzel {et~al.}(2021)Heinzel, Coughlin, Dietrich, Bulla, Antier,
  Christensen, Coulter, Foley, Issa, \& Khetan}]{Heinzel:2020qlt}
Heinzel, J., Coughlin, M.~W., Dietrich, T., {et~al.} 2021, Mon. Not. Roy.
  Astron. Soc., 502, 3057, \dodoi{10.1093/mnras/stab221}

\bibitem[{Hilditch {et~al.}(2013)Hilditch, Bernuzzi, Thierfelder, Cao, Tichy,
  {et~al.}}]{Hilditch:2012fp}
Hilditch, D., Bernuzzi, S., Thierfelder, M., {et~al.} 2013, Phys. Rev., D88,
  084057, \dodoi{10.1103/PhysRevD.88.084057}

\bibitem[{Huth {et~al.}(2022)Huth, Pang, Tews, Dietrich, Le~Fevre, Schwenk,
  Trautmann, Agarwal, Bulla, Coughlin, \& Van Den~Broeck}]{Huth:2021bsp}
Huth, S., Pang, P. T.~H., Tews, I., {et~al.} 2022, Nature, 606, 276,
  \dodoi{10.1038/s41586-022-04750-w}

\bibitem[{Komoltsev \& Kurkela(2022)}]{Komoltsev:2021jzg}
Komoltsev, O., \& Kurkela, A. 2022, Phys. Rev. Lett., 128, 202701,
  \dodoi{10.1103/PhysRevLett.128.202701}

\bibitem[{Lackey {et~al.}(2006)Lackey, Nayyar, \& Owen}]{Lackey:2005tk}
Lackey, B.~D., Nayyar, M., \& Owen, B.~J. 2006, Phys. Rev. D, 73, 024021,
  \dodoi{10.1103/PhysRevD.73.024021}

\bibitem[{Landry \& Essick(2019)}]{Landry:2018prl}
Landry, P., \& Essick, R. 2019, Phys. Rev. D, 99, 084049,
  \dodoi{10.1103/PhysRevD.99.084049}

\bibitem[{Margalit \& Metzger(2017)}]{Margalit:2017dij}
Margalit, B., \& Metzger, B.~D. 2017, Astrophys. J. Lett., 850, L19,
  \dodoi{10.3847/2041-8213/aa991c}

\bibitem[{Marti {et~al.}(1991)Marti, Ibanez, \& Miralles}]{Marti:1991wi}
Marti, J.~M., Ibanez, J.~M., \& Miralles, J.~A. 1991, Phys. Rev., D43, 3794,
  \dodoi{10.1103/PhysRevD.43.3794}

\bibitem[{Miller \& others (NICER~collaboration)(2019)}]{Miller:2019cac}
Miller, M.~C., \& others (NICER~collaboration). 2019, Astrophys. J. Lett., 887,
  L24, \dodoi{10.3847/2041-8213/ab50c5}

\bibitem[{Miller \& others (NICER~collaboration)(2021)}]{Miller:2021qha}
---. 2021, Astrophys. J. Lett., 918, L28, \dodoi{10.3847/2041-8213/ac089b}

\bibitem[{Most {et~al.}(2018{\natexlab{a}})Most, Papenfort, Dexheimer,
  Hanauske, Schramm, Stöcker, \& Rezzolla}]{Most:2018eaw}
Most, E.~R., Papenfort, L.~J., Dexheimer, V., {et~al.} 2018{\natexlab{a}},
  arXiv: 1807.03684

\bibitem[{Most {et~al.}(2018{\natexlab{b}})Most, Weih, Rezzolla, \&
  Schaffner-Bielich}]{Most:2018hfd}
Most, E.~R., Weih, L.~R., Rezzolla, L., \& Schaffner-Bielich, J.
  2018{\natexlab{b}}, Phys. Rev. Lett., 120, 261103,
  \dodoi{10.1103/PhysRevLett.120.261103}

\bibitem[{Mushotzky {et~al.}(2019)}]{Mushotzky:2019lpm}
Mushotzky, R.~F., {et~al.} 2019, Bull. Am. Astron. Soc., 51, 107.
\newblock \doarXiv{1903.04083}

\bibitem[{M\"uther {et~al.}(1987)M\"uther, Prakash, \&
  Ainsworth}]{Muther:1987xaa}
M\"uther, H., Prakash, M., \& Ainsworth, T.~L. 1987, Phys. Lett. B, 199, 469,
  \dodoi{10.1016/0370-2693(87)91611-X}

\bibitem[{Pang {et~al.}(2022)}]{Pang:2022rzc}
Pang, P. T.~H., {et~al.} 2022.
\newblock \doarXiv{2205.08513}

\bibitem[{Punturo {et~al.}(2010)Punturo, Abernathy, Acernese, Allen, Andersson,
  {et~al.}}]{Punturo:2010zz}
Punturo, M., Abernathy, M., Acernese, F., {et~al.} 2010, Class.Quant.Grav., 27,
  194002, \dodoi{10.1088/0264-9381/27/19/194002}

\bibitem[{Raaijmakers {et~al.}(2020)}]{Raaijmakers:2019dks}
Raaijmakers, G., {et~al.} 2020, Astrophys. J. Lett., 893, L21,
  \dodoi{10.3847/2041-8213/ab822f}

\bibitem[{Raaijmakers {et~al.}(2021)Raaijmakers, Greif, Hebeler, Hinderer,
  Nissanke, Schwenk, Riley, Watts, Lattimer, \& Ho}]{Raaijmakers:2021uju}
Raaijmakers, G., Greif, S.~K., Hebeler, K., {et~al.} 2021, Astrophys. J. Lett.,
  918, L29, \dodoi{10.3847/2041-8213/ac089a}

\bibitem[{Radice \& Dai(2019)}]{Radice:2018ozg}
Radice, D., \& Dai, L. 2019, Eur. Phys. J., A55, 50,
  \dodoi{10.1140/epja/i2019-12716-4}

\bibitem[{Read {et~al.}(2009)Read, Lackey, Owen, \& Friedman}]{Read:2008iy}
Read, J.~S., Lackey, B.~D., Owen, B.~J., \& Friedman, J.~L. 2009, Phys. Rev.,
  D79, 124032, \dodoi{10.1103/PhysRevD.79.124032}

\bibitem[{Reisswig \& Pollney(2011)}]{Reisswig:2010di}
Reisswig, C., \& Pollney, D. 2011, Class.Quant.Grav., 28, 195015,
  \dodoi{10.1088/0264-9381/28/19/195015}

\bibitem[{Reitze {et~al.}(2019)}]{Reitze:2019iox}
Reitze, D., {et~al.} 2019, Bull. Am. Astron. Soc., 51, 035.
\newblock \doarXiv{1907.04833}

\bibitem[{Rezzolla \& Takami(2016)}]{Rezzolla:2016nxn}
Rezzolla, L., \& Takami, K. 2016, Phys. Rev., D93, 124051,
  \dodoi{10.1103/PhysRevD.93.124051}

\bibitem[{Riley \& others (NICER~collaboration)(2019)}]{Riley:2019yda}
Riley, T.~E., \& others (NICER~collaboration). 2019, Astrophys. J. Lett., 887,
  L21, \dodoi{10.3847/2041-8213/ab481c}

\bibitem[{Riley \& others (NICER~collaboration)(2021)}]{Riley:2021pdl}
---. 2021, Astrophys. J. Lett., 918, L27, \dodoi{10.3847/2041-8213/ac0a81}

\bibitem[{Ruiz {et~al.}(2018)Ruiz, Shapiro, \& Tsokaros}]{Ruiz:2017due}
Ruiz, M., Shapiro, S.~L., \& Tsokaros, A. 2018, Phys. Rev., D97, 021501,
  \dodoi{10.1103/PhysRevD.97.021501}

\bibitem[{Somasundaram {et~al.}(2023)Somasundaram, Tews, \&
  Margueron}]{Somasundaram:2022ztm}
Somasundaram, R., Tews, I., \& Margueron, J. 2023, Phys. Rev. C, 107, L052801,
  \dodoi{10.1103/PhysRevC.107.L052801}

\bibitem[{Thierfelder {et~al.}(2011{\natexlab{a}})Thierfelder, Bernuzzi, \&
  Br{\"u}gmann}]{Thierfelder:2011yi}
Thierfelder, M., Bernuzzi, S., \& Br{\"u}gmann, B. 2011{\natexlab{a}},
  Phys.Rev., D84, 044012, \dodoi{10.1103/PhysRevD.84.044012}

\bibitem[{Thierfelder {et~al.}(2011{\natexlab{b}})Thierfelder, Bernuzzi,
  Hilditch, Br{\"u}gmann, \& Rezzolla}]{Thierfelder:2010dv}
Thierfelder, M., Bernuzzi, S., Hilditch, D., Br{\"u}gmann, B., \& Rezzolla, L.
  2011{\natexlab{b}}, Phys.Rev., D83, 064022,
  \dodoi{10.1103/PhysRevD.83.064022}

\bibitem[{Tichy(2009)}]{Tichy:2009yr}
Tichy, W. 2009, Class.Quant.Grav., 26, 175018,
  \dodoi{10.1088/0264-9381/26/17/175018}

\bibitem[{Tichy(2012)}]{Tichy:2012rp}
---. 2012, Phys. Rev. D, 86, 064024, \dodoi{10.1103/PhysRevD.86.064024}

\bibitem[{Tichy {et~al.}(2019)Tichy, Rashti, Dietrich, Dudi, \&
  Brügmann}]{Tichy:2019ouu}
Tichy, W., Rashti, A., Dietrich, T., Dudi, R., \& Brügmann, B. 2019, Phys.
  Rev. D, 100, 124046, \dodoi{10.1103/PhysRevD.100.124046}

\bibitem[{van Meter {et~al.}(2006)van Meter, Baker, Koppitz, \&
  Choi}]{vanMeter:2006vi}
van Meter, J.~R., Baker, J.~G., Koppitz, M., \& Choi, D.-I. 2006, Phys. Rev.,
  D73, 124011, \dodoi{10.1103/PhysRevD.73.124011}

\bibitem[{York(1999)}]{York:1998hy}
York, James~W., J. 1999, Phys.Rev.Lett., 82, 1350,
  \dodoi{10.1103/PhysRevLett.82.1350}

\bibitem[{Zhang {et~al.}(2019)}]{eXTP:2018anb}
Zhang, S.-N., {et~al.} 2019, Sci. China Phys. Mech. Astron., 62, 29502,
  \dodoi{10.1007/s11433-018-9309-2}

\end{thebibliography}

\end{document}